\documentclass[12pt]{iopart}

\usepackage{graphicx}
\usepackage{cite}
\usepackage{iopams}

\usepackage{amstext}

\begin{document}

\title[Filter-design perspective applying to dynamical decoupling of a multi-qubit system]{Filter-design perspective applying to dynamical decoupling of a multi-qubit system}

\author{Su Zhi-Kun and Jiang Shao-Ji}

\address{State Key Laboratory of Optoelectronic Materials and Technologies, Sun Yat-sen University, Guangzhou 510275, People's Republic of China}
\ead{stsjsj@mail.sysu.edu.cn}

\begin{abstract}
We employ the filter-design perspective and derive the filter functions according to nested Uhrig dynamical decoupling (NUDD) and Symmetric dynamical decoupling
(SDD) in the pure-dephasing spin-boson model with N qubits. The performances of
NUDD and SDD are discussed in detail for a two-qubit system. The analysis shows that (i) SDD outperforms NUDD for the bath with a
soft cutoff while NUDD approaches SDD as the cutoff becomes harder; (ii) if
the qubits are coupled to a common reservoir, SDD helps to protect the
decoherence-free subspace while NUDD destroys it; (iii) when the imperfect
control pulses with finite width are considered, NUDD is affected in both the
high-fidelity regime and coherence time regime while SDD is affected
in the coherence time regime only.

\end{abstract}

\maketitle

\section{Introduction}

Decoherence of quantum system is inevitable, since coupling of a system to its
surrounding environment exists all the time. Suppressing decoherence is
significantly important for a variety of fascinating quantum-information
tasks. In particular, protecting generated entangled states from decoherence is an important issue that has been addressed in different frameworks\cite{PhysRevA.74.052317,open.systems.information.dynamics.13.463,PhysRevA.79.052308,PhysRevA.78.060302,advanced.science.letters.2.459}.One of the most effective techniques to combat with the decoherence
process is dynamical decoupling (DD)\cite{PhysRevA.58.2733,PhysRevLett.82.2417}, which evolves from the Hahn echo\cite{PhysRev.80.580} and
develops for refocusing techniques in nuclear magnetic resonance (NMR)\cite{PhysRevLett.25.218,Mehring1983,Haeberlen1976}. The
central idea of DD is first explicitly introduced to protect qubit coherence
in\cite{PhysRevA.58.2733}, and it is soon incorporated within a general dynamical
symmetrization framework\cite{PhysRevLett.82.2417}. DD schemes operate by subjecting the
system of interest to suitable sequences of external control operations, with
the purpose of removing or modifying unwanted contributions to the underlying
Hamiltonian. Several DD pulse sequences, such as Carr-Purcell-Meiboom-Gill
(CPMG)\cite{PhysRev.94.630,RSI.29.688}, concatenated dynamical decoupling (CDD)\cite{PhysRevLett.95.180501,PhysRevA.75.062310,jpb.44.154003} and Uhrig dynamical decoupling
(UDD)\cite{PhysRevLett.98.100504}, have been proposed by now. The above DD sequences focus on
single-qubit systems, but some studies have been extended to multi-qubit
systems recently\cite{PhysRevA.76.042310,PhysRevA.81.012331,PhysRevA.82.052338,PhysRevA.83.022306}. Remarkably, in\cite{PhysRevA.83.022306} a good mathematical understanding of nested UDD
(NUDD) is presented to suppress decoherence in arbitrary N-qubit system. NUDD scheme is based on mutually orthogonal operation set (MOOS).
In addition, there exist more general DD schemes to protect a multi-qubit
system, i.e. the periodic DD (PDD) and the symmetric DD (SDD)\cite{PhysRevLett.82.2417}. Both PDD and SDD
are based on an appropriate DD group, and the two schems can be used to
eliminate errors to the first and the second order in the Magnus expansion, respectively.

In this paper, we aim to provide some guidelines in the choice of sequences to
be applied in experiments on the suppression of decoherence for a multi-qubit
system. To realize the goal, firstly we employ the concept of DD
pulse-sequence construction as a filter-design problem\cite{jpb.44.154002} and derive the filter
functions according to NUDD and SDD in the pure-dephasing spin-boson model
with $N$ qubits. The $N$ qubits embedded in two limiting cases, a common
environment and separated environments, are considered. Furthermore, we discuss
in detail the performances of NUDD and SDD for two qubits. In the case of
a single qubit, it is shown that the UDD sequence outperforms the CPMG sequence
for pure dephasing noises with sharp high-frequency cutoff while it performs slightly worse for
soft cutoffs\cite{njp.10.083024,PhysRevLett.100.160505,Nature.458.996,PhysRevA.79.062324,PhysRevLett.103.040501}. As for a two-qubit system, it is also significant to discuss the
performances of NUDD and SDD and see which performs better under the
soft-cutoff and hard-cutoff baths. How to simulate the soft-cutoff and
hard-cutoff baths? In terms of an experimental point, the noise used to
examine the DD sequences, such as classical noise and ohmic spectrum, is in
general governed by a power law spectrum $1/\omega^{\alpha}$ (hereafter, $\alpha>0$)\cite{PhysRevA.81.012309}. Varying the exponent of the power law of the spectrum
allows us to simulate both baths with soft and hard cutoffs. On the other hand, how to test the performance of DD sequence? Exactly, a recent
study provides us a novel way, which is determined by the sequence itself
only, to measure the performance on the suppression of decoherence when the
system is subject to the noise scaling such as $\propto1/\omega^{\alpha}$\cite{PhysRevA.81.012309}. The result shows that,
except for the case of some small total number of pulses, SDD performs better
than NUDD for the bath with the $1/\omega$ (soft-cutoff) noise and NUDD almost
coincide with SDD but is still outperformed by SDD for the bath with the
$1/\omega^{4}$ (hard-cutoff) noise. In addition, SDD helps to save the
decoherence-free subspace while NUDD damages it. As the total pulse number
increases, all filter functions for the same DD sequence have the same
performance gradually in the two limiting cases mentioned above.

Then a new analytical framework of filter-design perspective is employed to
analyze some of the results observed\cite{jpb.44.154002}. The filter functions for NUDD and SDD in
the coherence time regime and high-fidelity regime are compared, respectively. The definition of the two regimes and the differences between them are given in \cite{jpb.44.154002}. Both high-fidelity and coherence time regime are ranges of frequency for a certain filter function. Filter functions are designed to increase the suppression of errors in high-fidelity regime at short times, while the error probabilities of a few tens of percent are accepted in coherence time regime. From the comparison, we find that the function in the coherence time
regime\ for SDD with the same total pulse number as NUDD peaks towards higher
frequencies than that of NUDD, which is similar to the difference between CPMG
and UDD. The functions in this regime can be used to demonstrate why SDD
performs better than NUDD. In addition, the filter function for SDD in the
high-fidelity regime has the same roffoff for various values of total pulse number while NUDD has different roffoffs for different values of total pulse number. The control
pulses in the studies now are assumed to be ideal pulses, which is unphysical.
Therefore, the control pulses with finite-width pulses are also considered in
this paper. We find that the finite width affects the coherence time regime only for SDD sequence while it has influence on both coherence time regime and high-fidelity regime for NUDD sequences.

The paper is organized as follows. In section 2, the dynamical decay of N
qubits under dephasing is analyzed and its evolution operator with general DD
strategy, SDD or NUDD, is derived in terms of a filter. In section 3, a two-qubit system, as
an example, is disscussed in detail and the result about it is demonstrated by
the filter-design perspective. The conclusion will be given in section 4.

\section{Decoherence suppression in a multi-qubit system}
\subsection{Dynamical decay under dephasing}
We consider $N$ qubits which do not interact directly among each other, while
each qubit may have a different coupling to the bath modes\cite{Hornberger.2009,PhysRevA.80.032314}. The total system
Hamiltonian, in units of $\hbar$, reads%

\begin{equation}
H=H_{S}+H_{B}+H_{SB},\label{eq.totalhami}%
\end{equation}
with
\begin{equation}
H_{S}=\sum_{j=0}^{N-1}\frac{\omega_{a}^{(j)}}{2}\sigma_{z}^{(j)}%
,\label{eq.systemhami}%
\end{equation}

\begin{equation}
\label{eq.bathhami}H_{B}=\sum_{k}\omega_{k}b_{k}^{\dag}b_{k}\ ,
\end{equation}

\begin{equation}
H_{SB}=\sum_{j=0}^{N-1}\sigma_{z}^{(j)}\sum_{k}\Big(q_{k}^{(j)}b_{k}^{\dag
}+(q_{k}^{(j)})^{\ast}b_{k}\Big),\label{interactionhami}%
\end{equation}
where the first and second contribution $H_{S}$ and $H_{B}$ describe,
respectively, the free evolution of the qubits and the environment, the third
term $H_{SB}$ describes a bilinear interaction between the two. $\omega
_{a}^{(j)}$ and $\sigma_{z}^{(j)}$ are the transition frequency and the
inversion operators of the $j$th qubit, $b_{k}^{\dagger}$ and $b_{k}$ are
creation and annihilation operators for the $k$th field mode, which are
characterized by a set of parameters $\{q_{k}^{(j)},\omega_{k}\}$. This
information is encoded in the spectral density%

\begin{equation}
J(\omega)=\sum_{k}\left\vert q_{k}^{(j)}\right\vert ^{2}\delta(\omega
-\omega_{k}).\label{spectral density}%
\end{equation}

The Hamiltonian $H_{SB}$ in Eq.(\ref{interactionhami}), in interaction picture
with respect to the free dynamics $(H_{S}+H_{B})$, is given as%
\begin{equation}
\widetilde{H}_{SB}(t)=\sum_{j=0}^{N-1}\sigma_{z}^{(j)}B^{(j)}(t),
\end{equation}
with%

\begin{equation}
B^{(j)}(t)=\sum_{k}\Big(q_{k}^{(j)}b_{k}^{\dag}\mbox{e}^{i\omega_{k}t}%
+(q_{k}^{(j)})^{\ast}b_{k}\mbox{e}^{-i\omega_{k}t}\Big).
\end{equation}

The evolution is determined by the time-ordered unitary operator%

\begin{equation}
\widetilde{U}_{f}(t)=\widehat{T}\exp\bigg\{-i\int_{0}^{t}ds\,\tilde{H}%
_{SB}(s)\bigg\}.\label{free unitary}%
\end{equation}

Under two standard assumptions: $(\text{i})$ the qubits and the environment are
initially uncorrelated and $(\text{ii})$ the environment is initially in
thermal equilibrium at temperature $T_{e}$,\ the reduced density matrix of the
multi-qubit system, in the standard product basis $\mathbb{B}=\{|0\rangle
,|1\rangle,\ldots,|N-1\rangle\}$ $($ an N-digit binary notation, e.g.,
$|6\rangle\equiv|110_{2}\rangle\equiv|\uparrow\uparrow\downarrow\rangle\ )$,
can be written as%

\begin{equation}
\widetilde{\rho}_{S}(t)=\mbox{Tr}_{B}\Big(\widetilde{U}_{f}(t)\widetilde{\rho
}_{S}(0)\widetilde{\rho}_{B}(0)\widetilde{U}_{f}^{\dagger}(t)\Big).
\end{equation}

The element of this reduced density matrix is%

\begin{eqnarray}
\left\langle m\right\vert \widetilde{\rho}_{S}(t)|n\rangle &  =\widetilde
{\rho}_{mn}(t)\nonumber\\
&  =\widetilde{\rho}_{mn}(0)\mbox{Tr}_{B}\Big[V_{mn}^{-}(t)\widetilde{\rho
}_{B}(0)\Big(V_{mn}^{+}(t)\Big)^{\dagger}\Big]\nonumber\\
&  =\widetilde{\rho}_{mn}(0)\Big<\Big(V_{mn}^{+}(t)\Big)^{\dagger}V_{mn}%
^{-}(t)\Big>,
\end{eqnarray}
with%

\begin{equation}
V_{mn}^{\pm}(t)=\widehat{T}\exp\bigg\{\pm i\int_{0}^{t}ds\sum_{j=0}%
^{N-1}(m_{j}-n_{j})B^{(j)}(s)\bigg\},
\end{equation}
where $m_{j}\in\{0,1\}$ indicates the $j$th digit in the binary representation
of the number $m$, for example, the $0$th, $1$st and $2$nd digit in the binary
representation of number $6$, i.e.$|6\rangle=|110_{2}\rangle$, are
$6_{0}=0,6_{1}=1,6_{2}=1$, respectively.

Thus we obtain%

\begin{equation}
\widetilde{\rho}_{mn}(t)=\widetilde{\rho}_{mn}(0)\zeta_{mn}(t),
\end{equation}
with%

\begin{equation}
\zeta_{mn}(t)=\bigg<\widehat{T}\exp\bigg\{-2i\int_{0}^{t}ds\,\sum_{j=0}%
^{N-1}(m_{j}-n_{j})B^{(j)}(s)\bigg\}\bigg>.\label{evolution mn}%
\label{eq.cohfun}
\end{equation}

Note that if $m_{j}=n_{j}$ the coherence function (\ref{eq.cohfun}) is zero and then the diagonal elements are constant, as expected in a pure dephasing model.

So far, the evolution without pulses has been analyzed. In the following
subsection, we incorporate sequence of pulses into the dynamical evolution so
that the decoherence of the system can be suppressed.

\subsection{Filter function formalism}

To suppress the decoherence in the pure-dephasing spin-boson model, we can
apply a certain number $D_{j}$ of ideal ($\delta$-shaped)$\ \pi$ pulses about
the $x$ axis on the $j$th qubit at times $T_{d}^{(j)}$, for $d=1,\ldots,D_{j}%
$, during the time interval from $T_{0}^{(j)}\equiv0$ to $T_{D_{j}+1}%
^{(j)}\equiv T$. Note that the pulses for different qubits may be applied at
the same or different times (i.e., the pulse operators may be single-qubit or
multi-qubit operations). After the pulses are incorporated into the free
dynamical evolution, the Eq.(\ref{evolution mn}) will be modified into\cite{phys.scr.82}

\begin{equation}
\zeta_{mn}(T)=\Big<W_{mn}(T)\Big>,
\end{equation}
with%

\begin{equation}
W_{mn}(T)=\widehat{T}\exp\bigg\{-2i\int_{0}^{T}ds\,\Big[\sum_{j=0}^{N-1}%
(m_{j}-n_{j})B^{(j)}(s)F^{(j)}(s)\Big]\bigg\},
\end{equation}
\begin{equation}
F^{(j)}(t)=\sum_{d=0}^{D_{j}-1}(-1)^{d}\theta(t-T_{d}^{(j)})\theta
(T_{d+1}^{(j)}-t),\label{timing}%
\end{equation}
where the step function $\theta(t)$\ is equal to $1$ if $t>0$\ and $0$ if
$t<0$. Here we discuss two limiting cases\cite{Hornberger.2009}: $($i$)$ Qubits feel a common
reservoir, i.e. $q_{k}^{(j)}=q_{k}$ . In this extreme, the separations of the
qubits are small compared to the wave length\ of the field modes. $($ii$) $
Qubits see independent reservoirs, i.e. $q_{k}^{(j)}=q_{k_{j}}$. Contrast to
the former, in this case the qubits are so far apart from each other that each
field mode couples only to a single qubit. We use the P-representation for the thermal density matrix\cite{PhysRevLett.10.277,PhysRevLett.10.84} and obtain the expectation value of $W_{mn}$\ \ %

\begin{equation}
\zeta_{mn}(T)\doteq\exp\bigg\{-\int_{0}^{\infty}d\omega\frac{S(\omega)}%
{\omega^{2}}\,F_{mn}(\omega T)\bigg\},\label{new evolution mn}%
\end{equation}
where the equal sign with a dot above $($ i.e. $\doteq)$\ represents
$c$-number phase factors have been omitted, and the noise spectrum $S(\omega)$ is
related to the spectral density $J(\omega)$ in Eq.(\ref{spectral density})\ by%

\begin{equation}
S(\omega)=J(\omega)\coth\left(  \frac{\omega}{2T_{e}}\right)  ,
\end{equation}
and the filter function\cite{new.j.phys.10.083024,PhysRevB.77.174509} for $q_{k}^{(j)}=q_{k}$ reads%
\begin{equation}
F_{mn}^{c}(\omega T)=\left\vert \sum_{j=0}^{N-1}(m_{j}-n_{j})f^{(j)}(\omega
T)\right\vert ^{2},\label{filter function mnc}%
\end{equation}
while for $q_{k}^{(j)}=q_{k_{j}}$, we arrive at%

\begin{equation}
F_{mn}^{i}(\omega T)=\sum_{j=0}^{N-1}\left\vert (m_{j}-n_{j})\right\vert
\left\vert f^{(j)}(\omega T)\right\vert ^{2},\label{filter function mni}%
\end{equation}
with the sampling function \ %

\begin{equation}
f^{(j)}(\omega T)=-i\omega\int_{0}^{T}\mbox{e}^{-i\omega t}F^{(j)}%
(t)dt=1+(-1)^{D_{j}+1}\mbox{e}^{-i\omega T}+2\sum_{d=1}^{D_{j}}(-1)^{d}%
\mbox{e}^{-i\omega T_{d}^{(j)}},\label{sampling function}%
\end{equation}
which encapsulates actually all information about an arbitrary sequence. The comparison between filter function (\ref{filter function mnc})
and (\ref{filter function mni}) clearly shows that the sum in
Eq.(\ref{filter function mnc}) can be understood as the interference of the
sampling functions applied on different qubits, while the sum in
Eq.(\ref{filter function mni}) doesn't indicate this interference effect. A
question is what different performances DD schemes have in these two limiting
cases. This is one of the subjects of this paper. On the other hand, to
measure the performance of different DD schemes, an approach proposed by
Pasini et al\cite{PhysRevA.81.012309}. can be used. The approach is suitable for power spectrum
$S(\omega)=\frac{S_{0}}{\omega^{\alpha}}(\alpha>0)$. We can write down the construction as

\begin{equation}
\frac{S(\omega)}{\omega^{2}}=\frac{S_{0}}{\omega^{\alpha+2}}.
\end{equation}

Then the decoherence function in Eq.(\ref{new evolution mn}) gives%
\begin{equation}
\chi_{mn}(T)=\int_{0}^{\infty}d\omega\frac{S(\omega)}{\omega^{2}}%
\,F_{mn}(\omega T)=S_{0}\int_{0}^{\infty}d\omega\frac{1}{\omega^{\alpha+2}%
}\,F_{mn}(\omega T).\label{decoherence function}%
\end{equation}

Substituting $z=$ $\omega T$ in the Eq.(\ref{decoherence function}), we obtain
\begin{equation}
\chi_{mn}(T)=S_{0}T^{\alpha+1}I,\label{simple decoherence function}%
\end{equation}
with%
\begin{equation}
I:=\int_{0}^{\infty}\frac{F_{mn}(z)}{z^{\alpha+2}}dz.\label{factor}%
\end{equation}
The Eq.(\ref{simple decoherence function}) reveals that the time dependence of
$\chi_{mn}(T)$ is a simple power of $T$. Meanwhile, the factor $I$\ is a
quantity which expresses how well the pulse sequence protects a system against
unwanted evolution. Note that we can simulate both baths with soft and hard
cutoffs by varying the exponent of the power law of the spectrum. The smaller
exponent $\alpha$\ is, the softer the UV behavior of the power law spectrum
is. Vice versa, the larger exponent $\alpha$\ is, the harder the UV behavior
of the power law spectrum is. The famous $1/f$ noise corresponds to $\alpha
=1$\ in the notation. The case $S(\omega)\varpropto$\ $1/\omega^{4}$ is
experimentally relevant for ions in a Penning trap\cite{Nature.458.996,PhysRevA.79.062324}. This case corresponds to
$\alpha=4$ in the above notation.\ The second question we want to report is
which DD schemes perform better for softer and harder baths.

We next describe two kinds of DD schemes, SDD and NUDD, for the model we
consider and give their specific expressions of Eq.(\ref{sampling function}).
Then non-ideal control pulse with finite-width is also considered.

\subsubsection{\bigskip SDD sequence}

SDD schemes is a group-based DD\cite{PhysRevLett.82.2417}. The element in the group
$\{g_{i}\}_{i=0}^{\left\vert G\right\vert -1}$ with order $\left\vert
G\right\vert $, where $g_{0}=I_{s}$, is actually a base of linear unitary operators
and the Hamiltonian, such as $H$ in Eq.(\ref{eq.totalhami}), can be expressed
as a linear combination of these bases. The building block of a group-based DD
is realized by applying the pulses $p_{i}=g_{i+1}g_{i}^{\dagger}$\ separated
by the same time delay $\Delta t$. Thus, the time evolution after a control
cycle $T=\left\vert G\right\vert \Delta t$ can be written as%

\begin{eqnarray}
U(T)  & =\Big(g_{\left\vert G\right\vert -1}^{\dagger}\mbox{e}^{-iH\Delta
t}g_{\left\vert G\right\vert -1}\Big)\Big(g_{\left\vert G\right\vert
-2}^{\dagger}\mbox{e}^{-iH\Delta t}g_{\left\vert G\right\vert -2}%
\Big)\cdots\Big(g_{0}^{\dagger}\mbox{e}^{-iH\Delta t}g_{0}\Big)\nonumber\\
& =\widehat{T}\exp\bigg\{-i\int_{0}^{T}H_{g}(s)ds\bigg\},
\end{eqnarray}
where $H_{g}(s)\equiv$ $g_{i}^{\dagger}\mbox{e}^{-iH\Delta t}g_{i} $,
for $s\in\left(  i\Delta t,(i+1)\Delta t \right]  $. In the standard
time-dependent perturbation theory formalism, the propagator is expanded up to
the second order as%

\begin{equation}
U(T)=1+\sum_{i}h_{i}+\sum_{i>j}h_{i}h_{j}+\frac{1}{2}\sum_{i}h_{i}^{2}%
+O(T^{3}),\label{expanded terms}%
\end{equation}
where $h_{i}\equiv-i\Delta t H_{g}(s)$. We shall say that $k$th-order
decoupling is achieved if the first $k$ orders of the expansions of the propagator
commutes with an arbitrary element of the group $\{g_{i}\}_{i=0}^{\left\vert
G\right\vert -1}$, and the first non-commuting term arises from the
$(k+1)$th-order term. It was shown that\cite{PhysRevA.78.012355}, $U(T)$ in Eq.(\ref{expanded terms}) realizes
the first-order decoupling, since we have the commuting correlation $\left[
\sum_{j=0}^{\left\vert G\right\vert -1}h_{j},g_{i}\right]  =0$, with the
limiting $T\rightarrow0$ , for $i=0,1,\ldots,\left\vert G\right\vert -1,$
while to the second-order term we haven't this commuting correlation. This
group-based cyclic sequence is referred\ to as PDD. Furthermore,
we can obtain second-order decoupling via so-called SDD. The
cycle becomes twice as long as PDD, $T^{SDD}=2T,$ and the time evolution
operator is given by%
\begin{equation}
U^{SDD}(2T)=\overline{U}U=1+2\sum_{i=0}^{\left\vert G\right\vert -1}%
h_{i}+\frac{1}{2!}\left(  2\sum_{i=0}^{\left\vert G\right\vert -1}%
h_{i}\right)  ^{2}+O(T^{3})\nonumber
\end{equation}
where $\overline{U}\equiv\mbox{e}^{h_{0}}\mbox{e}^{h_{1}}\ldots
\mbox{e}^{h_{\left\vert G\right\vert -2}}\mbox{e}^{h_{\left\vert G\right\vert
-1}}$\ is mirror symmetric with $U.$ In this case first-order and all
even-order terms commute with $g_{i},$ which generally makes SDD better than PDD.
\bigskip As for an N-qubit system\cite{PhysRevLett.82.2417}, a possible choice of the group is
$G=\{I_{s},\sigma_{\beta}^{(j)}\}^{\otimes N}$, $\beta=x,y,z,$
$j=1,\ldots,N,$ $\left\vert G\right\vert =4^{N}$. However, if the relevant
system-bath coupling is known to be linear in single-qubit operators,
$H_{SB}=\sum_{j,\beta}$ $\sigma_{\beta}^{(j)}\otimes B_{\beta}^{(j)}$, one
has a simplified group, $G=\{I_{s},\otimes_{j=0}^{N-1}\sigma_{\beta}^{(j)}%
\}$, $\beta=x,y,z,$ $j=1,\ldots,N-1,$ $\left\vert G\right\vert =4$. In
particular, a purely decoherence coupling, like $H_{SB}$ in
Eq.(\ref{interactionhami}), the decoupling group can be further simplified
into $G=\{I_{s},\otimes_{j=0}^{N-1}\sigma_{x}^{(j)}\}$, $\left\vert
G\right\vert =2$. Therefore, the SDD scheme for the system considered in this
study is just to repeatedly apply the following pulse sequence:
(fXfX)(XfXf)=fXffXf, where f and X denote a ``pulse-free" period of a fixed
duration and the matrix $\otimes_{j=0}^{N-1}\sigma_{x}^{(j)}$, respectively.
Then the sampling function (\ref{sampling function}) can be specified as%

\begin{equation}
f_{SDD}^{(j)}(\omega T)=\frac{4i\mbox{e}^{\frac{i\omega T}{2}}\sin
(\frac{\omega T}{2})\sin^{2}(\frac{\omega T}{4D_{j}})}{\cos(\frac{\omega
T}{2D_{j}})}\label{sampling function sdd}%
\end{equation}
where $D_{j}$ stands for the number of pulses applied on the $j$th qubit and
is an even number, the same value for every qubit, that is, $D_{0}%
=D_{1}=\cdots=D_{N-1}$.

\subsubsection{NUDD sequence}
NUDD schemes is a MOOS-based DD\cite{PhysRevA.83.022306}. A MOOS is defined as a set of
operators which are unitary and Hermitian and have the property that each pair
of elements either commutes or anticommutes. We can eliminate the effects of
unwanted interactions between a system and its environment by the protection
of this MOOS. A choice of the MOOS for our pure dephasing model is
$\{\sigma_{x}^{(j)}\}_{j=0}^{N-1}\equiv\{\sigma_{x}^{(0)},\sigma_{x}%
^{(1)},\ldots,\sigma_{x}^{(N-1)}\}$.

Now we describe the construction of NUDD for protecting $\{\sigma_{x}%
^{(j)}\}_{j=0}^{N-1}$. The construction consists of $N$ levels of control, the
zeroth level, \ldots, the $(N-2)$th level and the $(N-1)$th level. First, in
the $(N-1)$th level (the outermost level), $L_{N-1}$ operators of $\sigma
_{x}^{(N-1)}$ are applied at UDD timing%

\begin{equation}%
\begin{array}
[c]{ccc}%
T_{l_{N-1}}=T\sin^{2}\frac{l_{N-1}\pi}{2L_{N-1}+2}, & for & l_{N-1}%
=1,\ldots,L_{N-1}%
\end{array}
,\label{T L-1 level}%
\end{equation}
between $T_{0}\equiv0$\ and $T_{L_{N-1}+1}\equiv T$. $L_{N-1}$ could be either
odd or even. Then the free evolution in each interval is substituted by
$L_{N-2}$ operators of $\sigma_{x}^{(N-2)}$ applied at%
\begin{eqnarray}
\fl
\begin{array}
[c]{ccc}%
T_{l_{N-1},l_{N-2}}=T_{l_{N-1}}+(T_{l_{N-1}+1}-T_{l_{N-1}})\sin^{2}%
\frac{l_{N-2}\pi}{2L_{N-2}+2}, & for & l_{N-2}=1,\ldots,L_{N-2}%
\end{array}
,\label{T L-2 level}%
\end{eqnarray}
in each interval between $T_{l_{N-1},0}\equiv T_{l_{N-1}}$\ and $T_{l_{N-1}%
,L_{N-2}+1}\equiv T_{l_{N-1}+1}$ with $L_{N-2}$\ being an even number. This is
the $(N-2)$th level. So on and so forth, the $n$th level of control is
constructed by applying $L_{n}$ operators of $\sigma_{x}^{(n)}$ in each
interval between $T_{l_{N-1},\ldots,l_{n+1},0}\equiv T_{l_{N-1},\ldots
,l_{n+2},l_{n+1}}$\ and $T_{l_{N-1},\ldots,l_{n+1},L_{n}+1}\equiv
T_{l_{N-1},\ldots,l_{n+2},l_{n+1}+1}$ at%

\begin{eqnarray}
\fl
\begin{array}
[c]{ccc}%
T_{l_{N-1},\ldots,l_{n+1},l_{n}}=T_{l_{N-1},\ldots,l_{n+1}}+(T_{l_{N-1}%
,\ldots,l_{n+2},l_{n+1}+1}-T_{l_{N-1},\ldots,l_{n+1}})\sin^{2}\frac{l_{n}\pi
}{2L_{n}+2},\\&for &l_{n}=1,\ldots,L_{n}
\end{array}
\end{eqnarray}
with $L_{n}$ being an even number. Thus, the sampling function
(\ref{sampling function}) for the $(N-1)$th, $(N-2)$th and $n$th level can be
specified as%

\begin{equation}
f_{NUDD}^{(N-1)}(\omega T)=\mbox{e}^{\frac{i\omega T}{2}}\sum_{l_{N-1}%
=-L_{N-1}-1}^{L_{N-1}}(-1)^{l_{N-1}}\mbox{e}^{\frac{-i\omega T}{2}\cos
(\frac{l_{N-1}\pi}{L_{N-1}+1})}\label{L-1 level}%
\end{equation}

\begin{eqnarray}
\fl
f_{NUDD}^{(N-2)}(\omega T)=\sum_{l_{N-1}=0}^{L_{N-1}}\mbox{e}^{\frac
{i\omega\tau_{l_{N-1}}}{2}}\sum_{l_{N-2}=-L_{N-2}-1}^{L_{N-2}%
}(-1)^{l_{N-2}}\mbox{e}^{\frac{-i\omega\tau_{l_{N-1}}}{2}\cos
(\frac{l_{N-2}\pi}{L_{N-2}+1})},\label{L-2 level}%
\end{eqnarray}

\begin{eqnarray}
\fl
f_{NUDD}^{(n)}(\omega T)=\sum_{l_{N-1}=0}^{L_{N-1}}\cdots\sum_{l_{n+1}%
=0}^{L_{n+1}}\mbox{e}^{\frac{i\omega\tau_{l_{N-1},\ldots,l_{n+1}}}{2}%
}\sum_{l_{n}=-L_{n}-1}^{L_{n}%
}(-1)^{l_{n}}\mbox{e}^{\frac{-i\omega\tau_{l_{N-1},\ldots,l_{n+1}}}%
{2}\cos(\frac{l_{n}\pi}{L_{n}+1})},\label{l level}%
\end{eqnarray}
where the evolution intervals $\tau_{l_{N-1}}=T_{l_{N-1}+1}-T_{l_{N-1}}$ and $\tau_{l_{N-1},\ldots,l_{n+1}}%
=\ T_{l_{N-1},\ldots,l_{n+1}+1}-T_{l_{N-1},\ldots,l_{n+1}}$. It
should be noted that the $n$th nesting level of NUDD can be protected to the
$L_{n}$th order, while the overall protection of the MOOS is up to the $L$th
order with $L=\min\{L_{N-1},L_{N-2},\cdots,L_{0}\}$. Athough NUDD scheme can
protect a multi-qubit system in pure dephasing bath to a higher decoupling order
than SDD, it remains a question whether NUDD offers improvement compared to
SDD using the same total number of pulses and the same total pulse sequence
duration $T$. We next discuss this question in detail. The reason for fixing the total time is that when $T$ is set to a certain value, SDD has the same pulse interval for every qubit while NUDD has different pulse intervals for different qubits, i.e. intervals for NUDD may be larger or smaller than that for SDD. Therefore, evaluating which sequence it might favor is not easy if $T$ is a fixed value, which ensures a relative fair comparison.
\subsubsection{Realistic pulses\bigskip}

The above theoretical studies are under the assumption that the duration pulse
time $\tau_{\pi}=0$. When the pulses are much shorter than the other
timescales of the system and the bath, it is a good approximation to treat
them as infinitely short. But this assumption is unphysical since not all real
system will satisfy the precondition. Here we consider a more realistic case,
$\tau_{\pi}\neq0$, and suppose that the interaction between system and bath is
negligible during the application of each pulse. Then the general
sampling\ function (\ref{sampling function}) incorporated a nonzero $\tau
_{\pi}$ is modified\ into\cite{Nature.458.996,PhysRevA.79.062324}

\bigskip%

\begin{equation}
f^{(j),r}(\omega T)=1+(-1)^{D_{j}+1}\mbox{e}^{-i\omega T}+2\sum_{d=1}^{D_{j}%
}(-1)^{d}\mbox{e}^{-i\omega T_{d}^{(j),r}}\cos(\frac{\omega\tau_{\pi}}{2}),
\end{equation}
where\bigskip\ $T_{d}^{(j),r}$ is the center time of the $d$th $\pi$ pulse.
Therefore, the sampling fuctions (\ref{sampling function sdd}) for SDD
sequences can be modified into%
\begin{equation}
f_{SDD}^{(j),r}=\cos(\frac{\omega\tau_{\pi}}{2})f_{SDD}^{(j)}+[1-\cos
(\frac{\omega\tau_{\pi}}{2})](1-\mbox{e}^{-i\omega T}),
\end{equation}
and the sampling functions (\ref{L-1 level})$\sim$(\ref{l level}) for NUDD can
be rewritten as
\begin{equation}
f_{NUDD}^{(N-1),r}=\cos(\frac{\omega\tau_{\pi}}{2})f_{NUDD}^{(N-1)}%
+[1-\cos(\frac{\omega\tau_{\pi}}{2})](1-\mbox{e}^{-i\omega T}),
\end{equation}%
\begin{equation}
f_{NUDD}^{(N-2),r}=\cos(\frac{\omega\tau_{\pi}}{2})f_{NUDD}^{(N-2)}%
+[1-\cos(\frac{\omega\tau_{\pi}}{2})]\sum_{l_{N-1}=0}^{L_{N-1}}%
(1-\mbox{e}^{-i\omega\tau_{l_{N-1}}}),
\end{equation}%
\begin{equation}
f_{NUDD}^{(n),r}=\cos(\frac{\omega\tau_{\pi}}{2})f_{NUDD}^{(n)}+[1-\cos
(\frac{\omega\tau_{\pi}}{2})]\sum_{l_{N-1}=0}^{L_{N-1}}\cdots\sum_{l_{n+1}%
=0}^{L_{n+1}}(1-\mbox{e}^{-i\omega\tau_{l_{N-1},\ldots,l_{n+1}}}),
\end{equation}
where $\tau_{l_{N-1}}$ and $\tau_{l_{N-1},\ldots,l_{n+1}}$, as
before, are defined as $\tau_{l_{N-1}}=T_{l_{N-1}+1}-T_{l_{N-1}}$ and $\tau_{l_{N-1},\ldots,l_{n+1}}%
=\ T_{l_{N-1},\ldots,l_{n+1}+1}-T_{l_{N-1},\ldots,l_{n+1}}$. In
this study, we also analyze\ whether the roles of $\tau_{\pi}$ for SDD and
NUDD are the same or different.
\section{Numerical results and discussion}
\bigskip\ For simplicity, we next restrict our attention to two-qubit system
initially prepared in an arbitrary X two-qubit state, with only diagonal and anti-diagonal elements, under the stand product basis $B=\{|1\rangle=|\uparrow\uparrow\rangle,|2\rangle=|\uparrow\downarrow\rangle,|3\rangle=|\downarrow\uparrow\rangle,|4\rangle=|\downarrow\downarrow\rangle\}$. The MOOS for NUDD
sequence is $\{\sigma_{x}^{(j)}\}_{j=0}^{1}\equiv\{\sigma_{x}^{(0)},\sigma
_{x}^{(1)}\}$. From Eq.(\ref{L-1 level})$\thicksim$(\ref{l level}), we can
write down the sampling functions of this situation as%

\begin{equation}
f_{NUDD}^{(1)}(\omega T)=\mbox{e}^{\frac{i\omega T}{2}}\sum_{l_{1}=-L_{1}%
-1}^{L_{1}}(-1)^{l_{1}}\mbox{e}^{\frac{-i\omega T}{2}\cos(\frac{l_{1}\pi
}{L_{1}+1})}%
\end{equation}
\bigskip%
\begin{equation}
f_{NUDD}^{(0)}(\omega T)=\sum_{l_{1}=0}^{L_{1}}\mbox{e}^{\frac{i\omega
\tau_{l_{1}}}{2}}\sum_{l_{0}=-L_{0}-1}^{L_{0}}(-1)^{l_{0}}\mbox{e}^{\frac{-i\omega\tau_{l_{1}}}{2}\cos(\frac{l_{0}\pi}{L_{0}+1})},
\end{equation}
with%

\begin{equation}
\tau_{l_{1}}=T_{l_{1}+1}-T_{l_{1}}%
\end{equation}
where $T_{l_{1}+1}\ $and $T_{l_{1}}$ can be obtained from
Eq.(\ref{T L-1 level}), and $L_{1}$\ is an even
number here. Accordingly, when the non-ideal pulses with finite width $\tau_{\pi}$
are considered, the sampling functions can be modified into%

\begin{equation}
f_{NUDD}^{(1),r}=\cos(\frac{\omega\tau_{\pi}}{2})f_{NUDD}^{(1)}+[1-\cos
(\frac{\omega\tau_{\pi}}{2})](1-\mbox{e}^{-i\omega T}),
\end{equation}

\begin{equation}
f_{NUDD}^{(0),r}=\cos(\frac{\omega\tau_{\pi}}{2})f_{NUDD}^{(0)}+[1-\cos
(\frac{\omega\tau_{\pi}}{2})]\sum_{l_{1}=0}^{L_{1}}(1-\mbox{e}^{-i\omega
\tau_{l_{1}}}).
\end{equation}
On the other hand, the decoupling group for SDD sequence is $G=\{I_{s}%
,\sigma_{x}^{(0)}\sigma_{x}^{(1)}\}$, $\left\vert G\right\vert =2$. The
sampling functions and the modified sampling functions read%

\begin{equation}
f_{SDD}^{(j)}(\omega T)=\frac{4i\mbox{e}^{\frac{i\omega T}{2}}\sin
(\frac{\omega T}{2})\sin^{2}(\frac{\omega T}{4D_{j}})}{\cos(\frac{\omega
T}{2D_{j}})}%
\end{equation}

\begin{equation}
f_{SDD}^{(j),r}=\cos(\frac{\omega\tau_{\pi}}{2})f_{SDD}^{(j)}+[1-\cos
(\frac{\omega\tau_{\pi}}{2})](1-\mbox{e}^{-i\omega T}),
\end{equation}
with $j=0,1$. Note that the pulse numbers $D_{0}\ $and $D_{1}$ are even
numbers and equivalent. From Eq.(\ref{filter function mnc})$\ $and
(\ref{filter function mni}), we have the filter functions for the case of a
common reservoir%
\begin{equation}
F_{14}^{c}(\omega T)=\left\vert f^{(0)}(\omega T)+f^{(1)}(\omega T)\right\vert
^{2},
\end{equation}

\begin{equation}
F_{23}^{c}(\omega T)=\left\vert f^{(0)}(\omega T)-f^{(1)}(\omega T)\right\vert
^{2},
\end{equation}
and the filter functions for the case of two separated reservoirs%

\begin{equation}
F_{14}^{i}(\omega T)=F_{23}^{i}(\omega T)=\left\vert f^{(0)}(\omega
T)\right\vert ^{2}+\left\vert f^{(1)}(\omega T)\right\vert ^{2}.
\end{equation}

\begin{figure}
[ptb]
\begin{center}
\includegraphics[
height=4.8196in,
width=6.282in
]%
{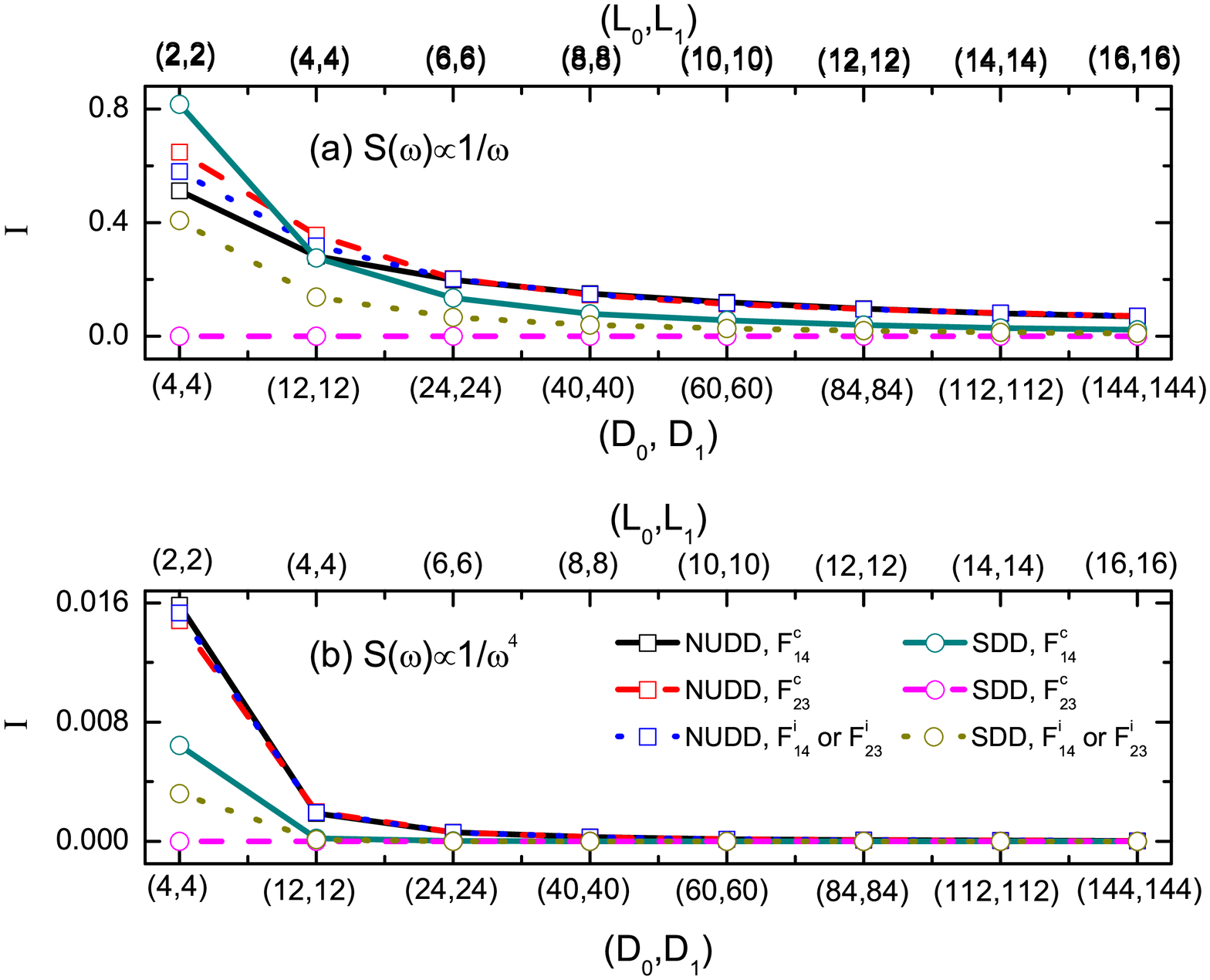}%
\caption{The factor $I$ as a function of the number of pulses $(L_{0},L_{1})$
for NUDD sequence (squares) and $(D_{0},D_{1})$ for SDD sequence (circles).
The pulse sequences are compared for different filter functions $F(z)$ under
the spectra (a) $S(\omega)\propto1/\omega$ and (b) $S(\omega)\propto
1/\omega^{4}$, see Eq.(\ref{factor}).}%
\label{fig1}%
\end{center}
\end{figure}

Now, the performances of NUDD and SDD sequences in the two limiting cases, a
common bath and two separated baths, are discussed.  To ensure a relative fair comparison, in this study we assume both
the total number of pulses and the total pulse sequence durations are the same
for these two schemes, i.e. $L_{0}+(L_{0}+1)L_{1}=D_{0}+D_{1}$ and
$T^{NUDD}=T^{SDD}$. In figure \ref{fig1}, the
factor $I$\cite{PhysRevA.81.012309}, see Eq.(\ref{factor}), is used to measure the performances of
NUDD and SDD sequences for different filter functions $F_{14}^{c}$,
$F_{23}^{c}$, $F_{14}^{i}$ and $F_{23}^{i}$ under the spectra $S(\omega
)\propto1/\omega$ and $S(\omega)\propto1/\omega^{4}$. The factors for
different filter functions with the same sequence have obviously different
performances if the number of pulses is small, while they almost coincide if
the number of pulses becomes larger. In addition, the evaluation of the factor
shows that SDD performs better than NUDD for the same filter function under
$S(\omega)\propto1/\omega$,\ while under $S(\omega)\propto1/\omega^{4}$ NUDD
performs similarly to SDD but is still outperformed by SDD over the range of
larger number of pulses. Note that there is a special dot $(D_{0}%
,D_{1})=(4,4)$ for the SDD sequence with filter function $F_{14}^{c}$, where
it performs abnormally and worse than NUDD for the same filter function. We
will come back to this point below. It is also interesting to notice that the
data for SDD with $F_{23}^{c}$ is always zero but it is not for NUDD sequence
with $F_{23}^{c}$. This means if the qubits are coupled to a common reservoir,
SDD helps to protect the decoherence-free subspace while NUDD destroys it.%

\begin{figure}
[ptb]
\begin{center}
\includegraphics[
height=4.8196in,
width=6.282in
]%
{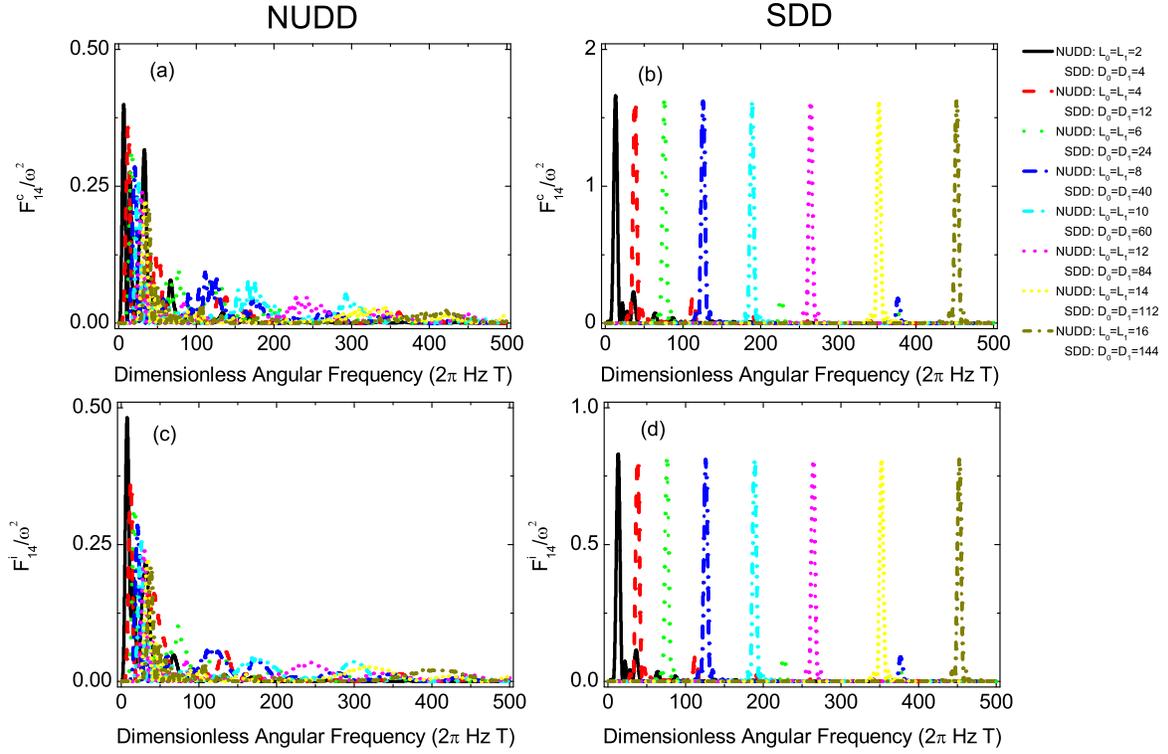}%
\caption{Modified filter functions, $F_{14}^{c}/\omega^{2}$ and $F_{14}%
^{i}/\omega^{2}$, as a function of dimensionless angular frequency $(\omega
T)$ \ for NUDD and SDD sequences with the total pulse numbers studied in
figure \ref{fig1}. This figure is related to the coherencce time regime and its layout is referred to reference\cite{jpb.44.154002}.}%
\label{fig2}%
\end{center}
\end{figure}

Then, we employ the newly proposed approach\cite{jpb.44.154002},including the figure layout, to discuss why SDD
performs better than NUDD and this question can be clarified in the coherence
time regime. The coherence time regime is a range of frequency for a filter function, in which the error probabilities of several tens of percent are accepted. The accumulated error leading to decoherence in this regime is very large. For clarity, we focus on the filter functions, $F_{14}^{c}$ and
$F_{14}^{i}$, and wrap the $1/\omega^{2}$ term into these functions, yielding
modified filter functions $F_{14}^{c}/\omega^{2}$ and $F_{14}^{i}/\omega^{2}$.
As shown in Eq.(\ref{decoherence function}), the decoherence function
$\chi(T)$ consists of the product of the spectrum $S(\omega)$ and the modified
filter function. The decoherence function is minimum if the overlap between
$S(\omega)$ and the modified filter function is minimum. The power law
spectrum $S(\omega)$ considered here is a strong low-frequency noise and the
intensity decreases as frequency increases. Smaller values of modified filter
functions in the low-frequency spectral regime will lead to smaller values of
decoherence function $\chi(T)$, and hence smaller factor $I$, supposing
$T^{NUDD}=T^{SDD}$. In figure \ref{fig2}, the filter functions as a function
of dimensionless angular frequency is\ shown for NUDD and SDD sequences with
the pulse numbers considered in figure \ref{fig1}. As expected, both NUDD and
SDD serve as high-pass filters and increasing total pulse number corresponds
to a shift in spectral peak towards higher frequencies. Thus figure \ref{fig1}
reveals that\ the factor $I$ decreases with pulse number increasing.
Furthermore, comparing figure \ref{fig2}(a) with figure \ref{fig2}(b) and
figure \ref{fig2}(c) with figure \ref{fig2}(d), we observe the modified filter
function for SDD peaks at higher frequencies than that for NUDD with the same
total pulse number when this number becomes larger. Therefore, SDD performs
better than NUDD when the pulse number is large, as shown in figure
\ref{fig1}. However, compared with SDD, the modified filter function
$F_{14}^{c}/\omega^{2}$ for NUDD peaks at almost the same frequencies but has a
smaller value when the pulse number is small, such as $(L_{0},L_{1})=(2,2)$ or
$(D_{0},D_{1})=(4,4)$. So it can be expected that NUDD performs better than
SDD for filter function $F_{14}^{c}$ with this small pulse number, as shown in
figure \ref{fig1}.%

\begin{figure}
[ptb]
\begin{center}
\includegraphics[
height=4.8196in,
width=6.282in
]%
{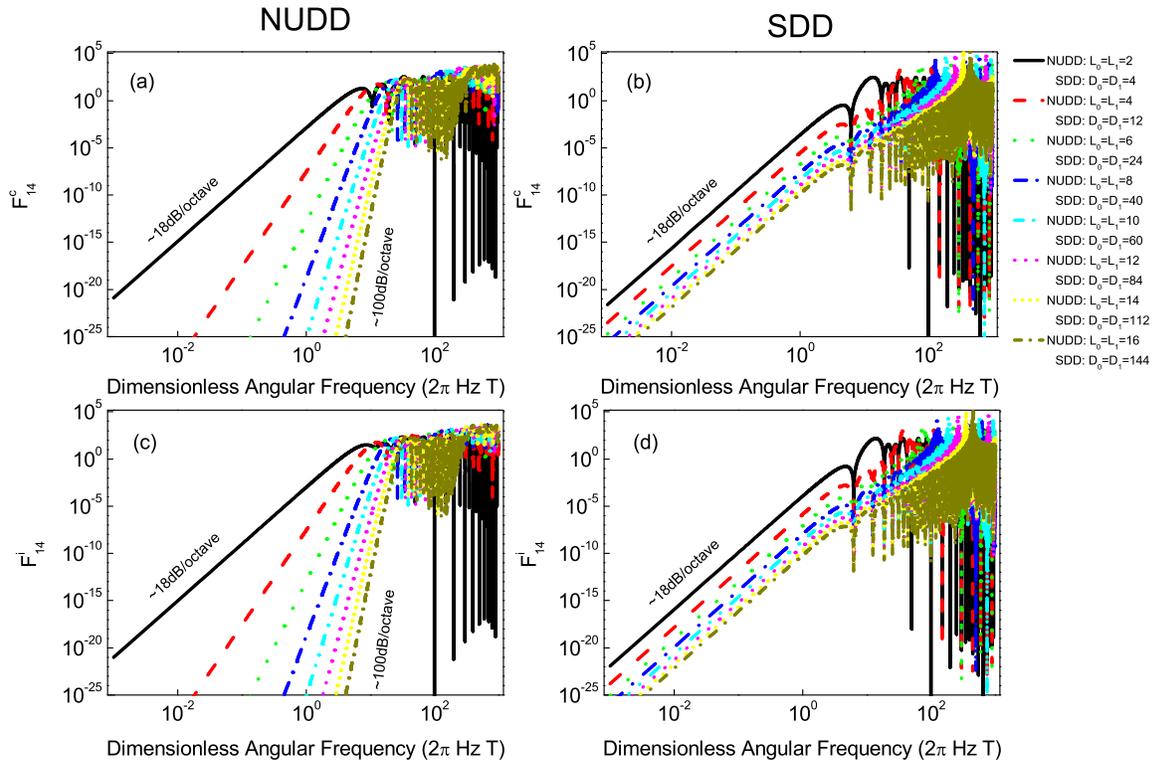}%
\caption{Log-Log plot of the filter functions, $F_{14}^{c}$ and $F_{14}^{i}$,
as a function of dimensionless angular frequency $(\omega T)$ \ for NUDD and
SDD sequences with the total pulse numbers studied in figure \ref{fig1}. This figure is related to the high-fidelity regime and its layout is referred to reference\cite{jpb.44.154002}.}%
\label{fig3}%
\end{center}
\end{figure}

Here, the comparison of the filter functions between NUDD and SDD in
another regime of interest, the high-fidelity regime, is also given. The high-fidelity regime, like the coherence time regime, is a range of frequency for a filter function, but in this regime the accumulated error bringing about decoherence is small. It turns up at short times $t \ll T_{2}$, where $T_{2}$ is the $1/e$ coherence time of a system. As before, we restrict our attention to the filter functions $F_{14}^{c}$ and
$F_{14}^{i}$. The filter functions for the high-fidelity regime are shown
graphically in figure \ref{fig3}, where the filter functions of NUDD and SDD
are presented on a log-log plot for various values of $(L_{0},L_{1})$ or
$(D_{0},D_{1})$. From figure \ref{fig3}(a) and \ref{fig3}(c), we find that
NUDD sequence provides a filter function whose low-frequency rolloff increases
from $\backsim18$dB/octave to $\backsim100$dB/octave as pulse number
increases from $(L_{0},L_{1})=(2,2)$ to $(L_{0},L_{1})=(16,16)$. By contrast,
the low-frequency rolloff for SDD is approximately constant, $\backsim
18$dB/octave, with each $(D_{0},D_{1})$. Here, rolloff is a term describing the steepness between the passband and the stopband of a filter function. The filter filters the noise in the stopband more efficiently, when the rolloff is steeper. Increasing rolloff means increasing the order of a filter and bringing the filter closer to the ideal response. If the rolloff is $6\eta dB/octave$, the order of the filter is considered as $\eta$. Then, as the total pulse number increases gradually, the filter order for NUDD becomes higher while the filter order for SDD stays about the same. Therefore, NUDD gives more flexibility and performs better than SDD, in terms of suppressing the low-frequency noise. Note that the performances of NUDD
and SDD are analogous to the performances of UDD and CPMG in the high-fidelity
regime for a single qubit, respectively\cite{jpb.44.154002}. The coherence time regime is
concerned with the decoherence function while the high-fidelity regime
provides small contributions to decoherence, but the high-fidelity regime
draws more stringent scrutiny than the coherence time regime, because the
high-fidelity regime is connected with predicted fault-tolerance error
thresholds of $p_{th}=0.01\%$ derived from quantum error correction and the
maximum allowable error must not surpass $p_{th}$ in quantum computing
applications\cite{jpb.44.154002}. The illustration about these two regimes here helps to understand
what changes when the pulses have nonzero $\tau_{\pi}$ in the following.%

\begin{figure}
[ptb]
\begin{center}
\includegraphics[
natheight=8.126600in,
natwidth=10.600000in,
height=4.8196in,
width=6.2777in
]%
{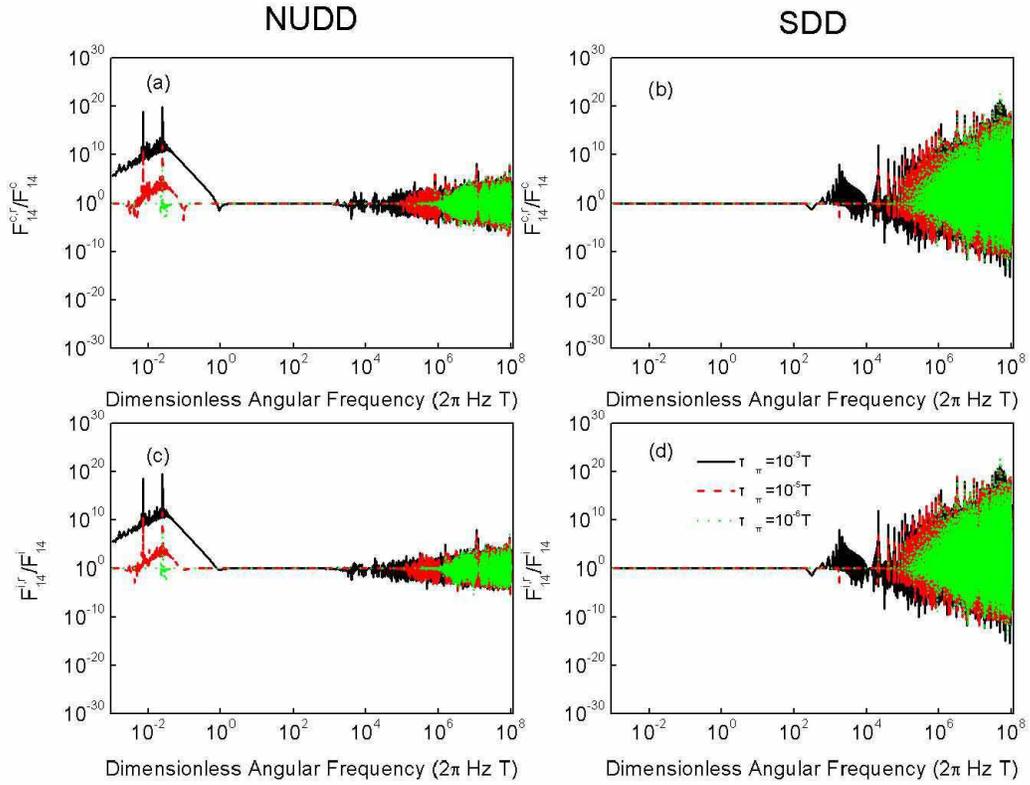}%
\caption{The ratio\ of the finite-width filter function $F_{14}^{c,r}$ (or
$F_{14}^{i,r}$) to the ideal filter function $F_{14}^{c}$(or $F_{14}^{i}$) for
NUDD with $(L_{0},L_{1})=(6,6)$ and SDD with $(D_{0},D_{1})=(24,24)$. In these
figures different traces correspond to different values of $\tau_{\pi}$ in
units of the total pulse duration time $T$.}%
\label{fig4}%
\end{center}
\end{figure}

Now let us compare the finite-width filter function with the ideal filter
function\cite{Nature.458.996,jpb.44.154002} to find out what is affected by the finite-width pulses. In figure
\ref{fig4}, we plot the ratio\ of the finite-width filter function
$F_{14}^{c,r}$ (or $F_{14}^{i,r}$) to the ideal filter function $F_{14}^{c}%
$(or $F_{14}^{i}$) for NUDD and SDD with a given total pulse number and here
we set $(L_{0},L_{1})=(6,6)$ or $(D_{0},D_{1})=(24,24)$. In these figures
different traces correspond to different values of $\tau_{\pi}$ in units of
the total pulse duration time $T$. From figure \ref{fig4}(a) and
\ref{fig4}(c), it can be seen that the filter function for NUDD is affected in
both coherence time regime and high-fidelity regime, still there is a mid
frequency range which is nearly unaffected. Moreover, as the pulse duration
$\tau_{\pi}$ increases, the influence on the filter function in the
high-fidelity regime is more remarkable, and the mid frequency range becomes
narrower. In comparison, figure \ref{fig4}(b) and \ref{fig4}(c) show that the
filter function for SDD is nearly unaffected in the high-fidelity regime while
the magnitude of the changes in coherence time regime is larger than that for
NUDD. Similarly, for SDD the range which is nearly unaffected becomes narrower
with the increasing of pulse duration $\tau_{\pi}$.%

Fig.4 is ploted numerically, in which the sampling step is $0.0001$
from $0.001$ to $10$ and $100$ from $10$ to $10^{8}$. In the following, more
details will be present. We first give out the analytical expression about the
ratio of the finite-width filter function to the ideal fiter function. For
convenience, we focus on the case of two independent reservoirs and the
expression for SDD reads

\begin{eqnarray}
\fl
R_{SDD}&=\frac{F_{14,SDD}^{i,r}}{F_{14,SDD}^{i}}\nonumber\\
\fl
& =\cos^{2}(\frac{\omega\tau_{\pi}}{2})+\frac{\sin^{2}(\frac{\omega\tau_{\pi}%
}{4})}{\sin^{2}(\frac{\omega T}{4D_{j}})}\cos(\frac{\omega T}{2D_{j}}%
)[\frac{\sin^{2}(\frac{\omega\tau_{\pi}}{4})}{\sin^{2}(\frac{\omega T}{4D_{j}%
})}\cos(\frac{\omega T}{2D_{j}})+2\cos(\frac{\omega\tau_{\pi}}{2})\cos(\omega
T)]\label{rsdd}%
\end{eqnarray}

It can be seen from Eq.(\ref{rsdd}) that there are singular points $\omega
T=4kD_{j}\pi$, with $k$ being integral number and $\sin^{2}(\frac{\omega
\tau_{\pi}}{4})\neq0$. On the other hand, the expression for NUDD is written as%

\begin{eqnarray}
\fl
R_{NUDD}  & =\frac{F_{14,NUDD}^{i,r}}{F_{14,NUDD}^{i}}\nonumber\\
\fl
& =\frac{(f_{Re}^{(0),r})^{2}+(f_{Im}%
^{(0),r})^{2}+(f_{Re}^{(1),r})^{2}+(f_{Im%
}^{(1),r})^{2}}{(f_{Re}^{(0)})^{2}+(f_{Im}%
^{(0)})^{2}+(f_{Re}^{(1)})^{2}+(f_{Im}^{(1)})^{2}}
\label{rnudd}%
\end{eqnarray}
where $f_{Re}^{(n)}$, $f_{Im}^{(n)}$ and
$f_{Re}^{(n),r} $, $f_{Im}^{(n),r}$ $(n=0,1)$
are real parts and image parts of the ideal filter function and of the
imperfect filter function, respectively. They are given by%

\begin{equation}
f_{NUDD}^{(n)}=f_{Re}^{(n)}+if_{Im}^{(n)}%
\end{equation}

\begin{equation}
f_{NUDD}^{(n),r}=f_{Re}^{(n),r}+if_{Im}^{(n),r}%
\end{equation}
with%

\begin{equation}
\fl
f_{Re}^{(1)}=\sum\limits_{l_{1}=-L_{1}-1}^{L_{1}}(-1)^{l_{1}%
}\{\cos(\frac{\omega T}{2})\cos[\frac{\omega T}{2}\cos(\frac{l_{1}\pi}%
{L_{1}+1})]+\sin(\frac{\omega T}{2})\sin[\frac{\omega T}{2}\cos(\frac{l_{1}%
\pi}{L_{1}+1})]\}
\end{equation}

\begin{equation}
\fl
f_{Im}^{(1)}=\sum\limits_{l_{1}=-L_{1}-1}^{L_{1}}(-1)^{l_{1}%
}\{-\cos(\frac{\omega T}{2})\sin[\frac{\omega T}{2}\cos(\frac{l_{1}\pi}%
{L_{1}+1})]+\sin(\frac{\omega T}{2})\cos[\frac{\omega T}{2}\cos(\frac{l_{1}%
\pi}{L_{1}+1})]\}
\end{equation}

\begin{equation}
\fl
f_{Re}^{(0)}=\sum\limits_{l_{1}=0}^{L_{1}}\sum\limits_{l_{0}%
=-L_{0}-1}^{L_{0}}(-1)^{l_{0}}\{\cos(\frac{\omega\tau_{l_{1}}}{2})\cos
[\frac{\omega\tau_{l_{1}}}{2}\cos(\frac{l_{0}\pi}{L_{0}+1})]+\sin(\frac
{\omega\tau_{l_{1}}}{2})\sin[\frac{\omega\tau_{l_{1}}}{2}\cos(\frac{l_{0}\pi
}{L_{0}+1})]\}
\end{equation}

\begin{equation}
\fl
f_{Im}^{(0)}=\sum\limits_{l_{1}=0}^{L_{1}}\sum\limits_{l_{0}%
=-L_{0}-1}^{L_{0}}(-1)^{l_{0}}\{-\cos(\frac{\omega\tau_{l_{1}}}{2})\sin
[\frac{\omega\tau_{l_{1}}}{2}\cos(\frac{l_{0}\pi}{L_{0}+1})]+\sin(\frac
{\omega\tau_{l_{1}}}{2})\cos[\frac{\omega\tau_{l_{1}}}{2}\cos(\frac{l_{0}\pi
}{L_{0}+1})]\}
\end{equation}

\begin{equation}
\fl
f_{Re}^{(1),r}=\cos(\frac{\omega\tau_{\pi}}{2}%
)f_{Re}^{(1)}+4\sin^{2}(\frac{\omega\tau_{\pi}}{4})\sin^{2}%
(\frac{\omega T}{2})
\end{equation}

\begin{equation}
\fl
f_{Im}^{(1),r}=\cos(\frac{\omega\tau_{\pi}}{2}%
)f_{Im}^{(1)}+2\sin^{2}(\frac{\omega\tau_{\pi}}{4})\sin(\omega T)
\end{equation}

\begin{equation}
\fl
f_{Re}^{(0),r}=\cos(\frac{\omega\tau_{\pi}}{2}%
)f_{Re}^{(0)}+4\sin^{2}(\frac{\omega\tau_{\pi}}{4})\sum
\limits_{l_{1}=0}^{L_{1}}\sin^{2}(\frac{\omega\tau_{l_{1}}}{2})
\end{equation}

\begin{equation}
\fl
f_{Im}^{(0),r}=\cos(\frac{\omega\tau_{\pi}}{2}%
)f_{Im}^{(0)}+2\sin^{2}(\frac{\omega\tau_{\pi}}{4})\sum
\limits_{l_{1}=0}^{L_{1}}\sin(\omega\tau_{l_{1}})
\end{equation}

It is a tough task to list the singular points of Eq.(\ref{rnudd}) one by one,
but we can study this problem numerically. Figure 5 is the enlarged view of figure 4(c) and (d) within three ranges, i.e. from $250$ to $1000$, from $9.99\times10^{5}$ to $1\times10^{6}$\ and\ from $3.999\times10^{6}$ to $4\times10^{6}$. From figure 5(a), we can see that the singular points appear periodically for the SDD sequency, as predicted by Eq.(\ref{rsdd}). However, figure 5(b) indicates that no periodical singular points are obtained within the ranges considered. Therefore,
the filter function for SDD is nearly unaffected in the high-fidelity regime,
that is, before the first singular point $(\omega T=4D_{j}\pi=96\pi\approx300, D_{j}=24)$ appears. In addition, SDD is affected by imperfect control pulses
in the coherence time regime, that is, after the first singular point $(\omega
T\approx300)$ presents. The numerical results also suggest that in the
coherence time regime the realistic filter function of NUDD is nearly
equivalent to the ideal filter function.%

\section{Conclusions}
In this paper we aim to provide some guidelines in the choice of sequences to be
applied in experiments on the suppression of decoherence for a multi-qubit system. We derived the filter functions for NUDD and SDD sequences with zero-width or finite-width pulses in the pure-dephasing spin-boson model. The filter design perspective was used to analyze the performances of these sequences. The performances of two qubits initially in X state were discussed in detail.

The analysis shows that SDD outperforms NUDD for the bath with a
soft cutoff while NUDD approaches SDD as the cutoff becomes harder. Second, if
the qubits are coupled to a common reservoir, SDD helps to protect the
decoherence-free subspace while NUDD destroys it. Third, when the imperfect
control pulses with finite width are considered, NUDD is affected in both the
high-fidelity regime and coherence time regime while SDD is obviously affected
in the coherence time regime only. Note that SDD scheme protects multiqubit systems using multi-qubit operations while NUDD do it using only single-qubit operations. Therefore, future work for SDD will analyze the effect of the precision, with which pulse location of each qubit in a multi-qubit operation is specified. In addition, the discussion for the case with more than two qubits will be given.

\section*{Acknowledgements}

This work is supported by the National Natural Science Foundation of China under Grant No.60977042 and the Guangdong Natural Science Foundation under Grant No.9151027501000070.

\section*{References}\label{refs}
\bibliographystyle{unsrt}
\bibliography{jpb}

\section*{List of figures}

\newpage

\begin{figure}
[ptb]
\begin{center}
\includegraphics[
height=2.5261in,
width=6.2898in
]%
{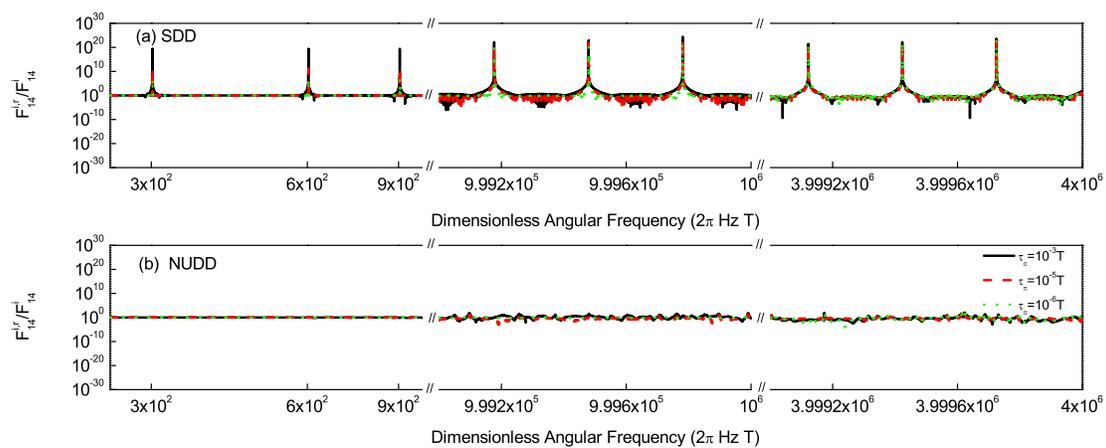}%
\caption{Enlarged view of figure 4(c) and (d) within the dimensionless
angular frequency ranges from $250$ to $1000 $, from $9.99\times10^{5}$ to
$1\times10^{6}$\ and\ from $3.999\times10^{6}$ to $4\times10^{6}$ for (a) SDD
sequence and (b) NUDD sequence. The parameters are the same as that of figure
4.}%
\label{fig5}%
\end{center}
\end{figure}

\label{lastpage}
\end{document}